\documentclass[%
 reprint,
 amsmath,amssymb,
 aps,
 prd,
 preprintnumbers,
nofootinbib,
]{revtex4-2}

\usepackage{graphicx}
\usepackage{dcolumn}
\usepackage{bm}
\usepackage[
colorlinks=true, 
citecolor=darkblue,  
linkcolor=darkred,   
urlcolor=darkblue
]{hyperref}

\usepackage[mathlines]{lineno}

\usepackage{xcolor}
\definecolor{darkred}{RGB}{200,0,0}
\definecolor{darkblue}{RGB}{40,80,220}
\usepackage{graphics,graphicx}

\definecolor{green}{rgb}{0.1,0.8,0.2}
\definecolor{orange}{rgb}{1.0,0.5,0.0}
\definecolor{cyan}{rgb}{0.0,0.75,0.8}
\definecolor{brown}{rgb}{0.7,0.35,0.05}

\newcommand{\GeV}{\, {\rm GeV}}
\newcommand{\ps}{\, {\rm ps}}
\newcommand{\psin}{\, {\rm ps^{-1}}}

\begin{document}

\preprint{TUM-HEP-1586/25, P3H-26-022, SI-HEP-2026-06, CERN-TH-2026-142}

\title{Predictions for \boldmath $b$-baryon lifetimes at NNLO-QCD}

\author{Alexander Lenz}
\email{alexander.lenz@uni-siegen.de}
\affiliation{Physik Department, Universit\"{a}t Siegen, Walter-Flex-Str.~3, D-57068 Siegen, Germany
\\
Cluster of Excellence “Color meets Flavor”}

\author{Maria Laura Piscopo}
\email{maria.laura.piscopo@cern.ch}

\affiliation{Theoretical Physics Department, CERN, 1211 Geneva 23, Switzerland}

\author{Aleksey~V.~Rusov}
\email{aleksey.rusov@tum.de}
\affiliation{Physik Department T31, 
Technische Universit\"at M\"unchen, James-Franck-Straße 1, D-85748 Garching, Germany}

\date{\today}

\begin{abstract}
Motivated by recent and forthcoming experimental progress,
we provide updated predictions for the total decay rates of $b$-baryons, their lifetime ratios and their lifetimes normalised to that of the $B^0_d$ meson within the framework of the heavy quark expansion (HQE).
We include, for the first time, next-to-next-to-leading-order QCD corrections to the free $b$-quark decay, which significantly reduce the theoretical uncertainties in the total decay rates. In addition, we also include, for the first time, the complete next-to-leading-order QCD corrections to the dimension-five contributions. While these corrections have only a minor effect on the total decay rates, they induce a sizeable shift in the lifetime ratios, improving the agreement between HQE predictions and experimental data. Overall, we find excellent agreement between the HQE predictions and current experimental measurements for both total decay rates and lifetime ratios within the quoted uncertainties.
\end{abstract}

\maketitle

\section{Introduction}
Lifetimes of hadrons containing a $b$ quark provide stringent tests of the heavy quark expansion (HQE) \cite{Khoze:1983yp,Shifman:1984wx,Blok:1992hw,Blok:1992he,Bigi:1991ir,Bigi:1992su,Beneke:1998sy}. Within this framework, the decay rate is expressed as a double expansion in inverse powers of the heavy-quark mass and in the strong coupling $\alpha_s$. The non-perturbative dynamics is encoded in hadronic matrix elements of local operators of increasing dimension, while the short-distance physics is described by Wilson coefficients that can be computed perturbatively in QCD.

The HQE has been successfully applied to the description of $B$-meson~\cite{Lenz:2022rbq, Egner:2024lay} and $b$-baryon~\cite{Gratrex:2023pfn,Dulibic:2026atz}  lifetimes, see the reviews~\cite{Lenz:2014jha,Albrecht:2024oyn}. It also provides a consistent description of charm-hadron lifetimes, reproducing the observed pattern for $D$ mesons~\cite{King:2021xqp} and charmed baryons~\cite{Gratrex:2022xpm}, albeit with larger theoretical uncertainties. In the bottom sector, the achieved precision allows lifetime observables to provide constraints on possible contributions from physics beyond the Standard Model, in particular through theoretically cleaner quantities such as lifetime ratios~\cite{Lenz:2022pgw, Lang:2025ios,Meiser:2024zea}. Continued improvement of HQE predictions for $b$-hadron lifetimes is therefore of central importance in testing both the HQE and the Standard Model.

In recent years, significant progress has been made in refining the HQE predictions. On the perturbative side, next-to-next-to-leading-order (NNLO) QCD corrections to the leading dimension-three contribution have been computed~\cite{Egner:2024azu}, together with next-to-leading-order~(NLO) QCD corrections to dimension-five two-quark operators~\cite{Mannel:2024uar,Mannel:2023zei,Mannel:2025fvj}, and, very recently, NNLO-QCD corrections to dimension-six four-quark operators in full QCD~\cite{Moretti:2026waq}. On the non-perturbative side, first Lattice QCD determinations of matrix elements of dimension-six four-quark operators based on the gradient-flow method have become available~\cite{Black:2026rbz,Black:2026dzp}, although these are currently restricted to charmed mesons.

The NNLO results for the partonic decay rate have already been incorporated in updated theoretical predictions of $B$-meson lifetimes~\cite{Egner:2024lay}, and, very recently, of doubly $b$-baryon and $B_c$ meson lifetimes~\cite{Dulibic:2026atz}, leading to a significant reduction of theoretical uncertainties. However, a comprehensive analysis, including these corrections is still missing for singly $b$-baryons. 
The state-of-the-art in this case can be found in Ref.~\cite{Gratrex:2023pfn}, which includes NLO-QCD corrections to both the partonic rate and the dimension-six four-quark operator contributions, and reaches only LO accuracy for the remaining power-suppressed contributions.

Experimentally, lifetimes of singly $b$ baryons have been measured with high precision~\cite{HFLAV, PDG:2026},
\footnote{Based on the measurements from Refs.~\cite{UA1:1991vse, ALEPH:1992yid, OPAL:1995nmi, DELPHI:1996eqs, DELPHI:1999con, ALEPH:1997ake, DELPHI:1995jet, DELPHI:2005zmk, CDF:1996fsp, D0:2007pfp, CDF:2009sse, CMS:2013bcs, CDF:2014mon, LHCb:2014wvs, LHCb:2014wqn, LHCb:2014jst, LHCb:2014chk, LHCb:2016coe, LHCb:2024lao, LHCb:2025zft}.} although generally with larger uncertainties than those of $B$ mesons. Among them, the $\Lambda_b^0$ lifetime is the most precisely determined, with sub-percent accuracy. The lifetimes of the $\Xi_b^-$ and $\Xi_b^0$ baryons have also been measured with uncertainties at the percent level. In contrast, the $\Omega_b^-$ lifetime remains the least well constrained, with a current experimental uncertainty of approximately $10\%$.

Recently, new measurements of the $\Xi_b^-$ and $\Xi_b^0$ lifetimes have been published and incorporated into the Heavy Flavour Averaging Group (HFLAV) averages. In addition, for the first time, the corresponding lifetime ratios with respect to the $\Lambda_b^0$ baryon have been determined. Specifically, the LHCb collaboration reported~\cite{LHCb:2025zft}
\begin{align}
\frac{\tau (\Xi_b^0)}{\tau (\Lambda_b^0)} & 
= 1.004 \pm 0.008 \pm 0.005\,, 
\end{align}
and \cite{LHCb:2024lao}
\begin{align}
\frac{\tau (\Xi_b^-)}{\tau (\Lambda_b^0)} & = 1.078  \pm 0.012  \pm 0.007\,. 
\end{align}

In light of the recent theoretical developments and the increasing experimental precision, in this work we present an updated HQE analysis of singly $b$-baryon lifetimes and of their ratios. Compared to Ref.~\cite{Gratrex:2023pfn}, we include, for the first time, NNLO-QCD corrections at dimension three and NLO-QCD corrections to the dimension-five two-quark operator contributions. Moreover, we update the numerical inputs for the matrix elements of dimension-five and dimension-six two-quark operators. These improvements substantially reduce the theoretical uncertainties, enabling more stringent tests of the HQE and more precise comparisons with the latest experimental measurements.

The remainder of this paper is organised as follows. In Section~\ref{sec:Updated_Analysis}, we review the current status of the HQE and describe the updates implemented in the present analysis. Our numerical results are presented in Section~\ref{sec:Results}, while our conclusions and outlook are given in Section~\ref{sec:Conclusion}.

\section{Updated analysis}
\label{sec:Updated_Analysis}
Within the HQE, the total decay width of a $b$ hadron can be schematically written as
\begin{eqnarray}
    \Gamma & = & \Gamma_3 + \Gamma_5 \frac{\langle {\cal O}_5 \rangle }{m_b^2} + \Gamma_6 \frac{\langle {\cal O}_6 \rangle }{m_b^3} + \ldots
    \nonumber
    \\
     &  & + 16 \pi^2 \left[  
     \tilde \Gamma_6 \frac{\langle \tilde{\cal O}_6 \rangle }{m_b^3} + 
     \tilde\Gamma_7 \frac{\langle \tilde{\cal O}_7 \rangle }{m_b^4} + \ldots \right]\,, 
     \label{eq:HQE}
\end{eqnarray}
where $\langle {\cal O}_i\rangle$ and $\langle \tilde{\cal O}_i\rangle$ denote hadronic matrix elements of local two- and four-quark operators of dimension $i$, while $\Gamma_i$ and $\tilde\Gamma_i$ are the corresponding Wilson coefficients. The latter can be computed perturbatively as an expansion in the strong coupling,
\begin{equation}
    \Gamma_i = \Gamma_i^{(0)} 
    + \frac{\alpha_s (\mu_b)}{4 \pi } \Gamma_i^{(1)}
    + \left( \frac{\alpha_s (\mu_b)}{4 \pi } \right)^2 \Gamma_i^{(2)}
     + \ldots \, ,
\end{equation}
evaluated at the renormalisation scale $\mu_b \sim m_b$. Details of the HQE for $b$ baryons, including complete expressions for the LO dimension-six matching coefficients, as well as the parametrisation of the dimension-six four-quark matrix elements, can be found in Ref.~\cite{Gratrex:2023pfn}. In the following, we discuss the corrections included in the present analysis and highlight the main improvements with respect to the previous study.

Starting with the perturbative contributions,\footnote{The perturbative accuracy of the short-distance coefficients is limited by the currently available results for non-leptonic $b$-quark decays. For consistency with previous analyses, we do not include higher-order QCD corrections that are currently known only for semileptonic decay modes.} the main improvements of this work are the inclusion of ${\cal O}(\alpha_s^2)$ corrections to the free $b$-quark decay rate and of ${\cal O}(\alpha_s)$ corrections to the contribution of the power-suppressed dimension-five two-quark operators, namely the kinetic and chromomagnetic operators.

For the partonic rate, the NNLO-QCD results in the case of semileptonic decays were obtained in Refs.~\cite{Fael:2020tow,Fael:2024gyw}, while those in the case of non-leptonic decays were recently computed in Ref.~\cite{Egner:2024azu}.
Throughout this work, we employ, as our default scenario, the kinetic scheme for the bottom-quark mass and the $\overline{\rm MS}$ scheme for the charm-quark mass. For the partonic contribution, this implies consistently expanding all expressions up to ${\cal O}(\alpha_s^2)$. The three-loop conversion formulae between the pole, kinetic, and $\overline{\rm MS}$ mass schemes are taken from Ref.~\cite{Fael:2020iea}, while the running and decoupling of the strong coupling and quark masses are implemented using \texttt{RunDec}~\cite{Herren:2017osy}. Overall, our treatment of the partonic contribution closely follows the recent NNLO analysis of $B$-meson lifetimes in Ref.~\cite{Egner:2024lay}, to which we refer for further details.

Turning to the inclusion of
NLO-QCD corrections to $1/m_b^2$-suppressed contributions, the case of the kinetic operator is straightforward, since reparametrisation invariance, see, e.g.,~Refs.~\cite{Dugan:1991ak, Chen:1993np, Luke:1992cs, Manohar:2010sf, Gunawardana:2017zix, Mannel:2018mqv}, implies that its Wilson coefficient is given by $-\Gamma_3/2$ to all orders in perturbation theory.

The situation is different for the chromomagnetic operator. Although the corresponding NLO expressions for semileptonic decays have been available for some time~\cite{Alberti:2013kxa,Mannel:2014xza,Mannel:2015jka,Mannel:2021zzr,Moreno:2022goo}, the ${\cal O}(\alpha_s)$ corrections to the matching coefficients induced by non-leptonic $b$-quark decays have only recently been computed. The NLO results for transitions with one massive quark in the final state, i.e.\ $b \to c\bar{u}d$, were obtained in Refs.~\cite{Mannel:2024uar,Mannel:2023zei}, while those for transitions with two equally massive quarks in the final state, i.e.\ $b \to c\bar{c}s$, were computed in Ref.~\cite{Mannel:2025fvj}. These corrections have been found to have a sizeable impact on the Wilson coefficient: they lift the accidental cancellations that affect the coefficient of the chromomagnetic operator at LO-QCD, see, e.g., Ref.~\cite{Lenz:2020oce}, and significantly reduce the renormalisation-scale dependence. Because these contributions are suppressed by $1/m_b^2$, their effect on the total lifetime is modest. However, they constitute an essential ingredient in lifetime ratios, where the dependence on the dominant partonic contribution cancels completely or is substantially reduced.

Concerning the remaining power corrections, our analysis closely follows that of Ref.~\cite{Gratrex:2023pfn}. Here, we briefly summarise the main ingredients. Starting at order $1/m_b^3$, both two- and four-quark operators contribute. For the dimension-six two-quark operators, we employ the LO QCD results of Refs.~\cite{Lenz:2020oce,Mannel:2020fts,Moreno:2020rmk,Rahimi:2022vlv,Moreno:2022goo}, since the corresponding ${\cal O}(\alpha_s)$ corrections induced by non-leptonic $b$-quark decays are not yet available, preventing a complete NLO analysis.
For the dimension-six four-quark operators, we use the NLO short-distance coefficients computed in Refs.~\cite{Franco:2002fc,Lenz:2013aua}, which correspond to operators defined in the heavy-quark effective theory (HQET) limit. Very recently, the first calculation of the $\mathcal{O}(\alpha_s^2)$ corrections to the dimension-six four-quark operator contributions was performed in Ref.~\cite{Moretti:2026waq} for operators defined in full QCD. However, because of the current difficulty in consistently accounting for the mixing of dimension-six four-quark operators into dimension-three two-quark operators under renormalisation in QCD, see, e.g., Refs.~\cite{Moretti:2026waq,Franco:2002fc,Beneke:2002rj}, these results can presently be applied only to lifetime ratios in which these mixing effects cancel, such as $\tau(B^+)/\tau(B_d^0)$ in the isospin limit or $\tau(D_s^+)/\tau(D^0)$ in the $V$-spin limit. For this reason, we do not include these corrections in the current analysis, as they could only be consistently incorporated in the ratio $\tau(\Xi_b^0)/\tau(\Xi_b^-)$.

Finally, at order $1/m_b^4$, complete results are not yet available even at LO in QCD. The matching coefficients of dimension-seven two-quark operators are only known in the case of semileptonic $b$ decays~\cite{Dassinger:2006md, Mannel:2023yqf},\footnote{Ref.~\cite{Mannel:2023yqf} also determined the contribution of dimension-eight two-quark operators from semileptonic $b \to c$ decays.} while for the contribution of four-quark operators, the Wilson coefficients are known at LO-QCD, however estimates of the corresponding hadronic matrix elements are still missing. Therefore, following Ref.~\cite{Gratrex:2023pfn}, we account for the effect of dimension-seven operators as an additional source of uncertainty.

\begin{table}[t]
\renewcommand{\arraystretch}{2.3}
\centering
\begin{tabular}{c|c}

\hline
$\displaystyle \langle {\cal O}_1^u \rangle_{\Lambda_b^0} = \langle {\cal O}_1^d \rangle_{\Lambda_b^0}$ & 
$ -0.0132 \pm 0.0022 \pm 0.0040 $ \\[0.0em]
\hline
$\displaystyle \langle {\cal O}_1^u \rangle_{\Xi_b^0} = \langle {\cal O}_1^d \rangle_{\Xi_b^-}$ & 
$ -0.0137 \pm 0.0024 \pm 0.0041 $ \\
\hline
$\displaystyle \langle {\cal O}_1^s \rangle_{\Xi_b^0} = \langle {\cal O}_1^s \rangle_{\Xi_b^-}$ & 
$-0.0179 \pm 0.0032 \pm 0.0054 $
\\
\hline
$\displaystyle \langle {\cal O}_1^s \rangle_{\Omega_b^-} $
& $-0.1270 \pm 0.0229 \pm 0.0381 $
\\
\hline
\end{tabular}
\caption{\small
Numerical values in GeV$^3$ of the baryonic matrix elements of the dimension-six colour-singlet four-quark operator ${\cal O}_1^q$, obtained within the NRCQM at the hadronic scale $\mu_h=1.5 \mbox{ GeV}$. The first uncertainties are obtained by varying all input parameters, while the second correspond to a conservative $30\%$ estimate of the model uncertainty.
} 
\label{tab:O1Numerics}
\end{table}

Having discussed the structure of the perturbative series entering our analysis, we now turn to the non-perturbative inputs, namely the hadronic matrix elements appearing in Eq.~\eqref{eq:HQE}.

It is convenient to begin with the dimension-six four-quark operators. We use the operator basis defined in terms of colour singlet and colour rearranged operators, where the $(V-A)\times(V-A)$ operators are given by
\begin{align}
{\cal O}^q_1 = \left(\bar h_v^i \gamma_\mu (1-\gamma_5) q^i\right) 
\left(\bar q^j \gamma^\mu (1-\gamma_5) h_v^j\right)\,,
\label{eq:O1q}
\\[2mm]
\tilde{\cal O}^q_1 = \left(\bar h_v^i \gamma_\mu (1-\gamma_5) q^j\right) 
\left(\bar q^j \gamma^\mu (1- \gamma_5) h_v^i\right)\,,
\end{align}
while the $(S-P)\times(S-P)$ operators are
\begin{align}
{\cal O}^q_2 = \left(\bar h_v^i  (1-\gamma_5) q^i\right) 
\left(\bar q^j (1+\gamma_5) h_v^j\right)\,,
\\[2mm]
\tilde {\cal O}^q_2 = \left(\bar h_v^i  (1-\gamma_5) q^j\right) 
\left(\bar q^j (1+\gamma_5) h_v^i\right)\,.
\label{eq:O2tq}
\end{align}
Here, $i$ and $j$ are colour indices and $h_v$ denotes the heavy-quark field in the HQET limit~\cite{Neubert:1993mb}. Baryonic matrix elements of the colour-singlet operators ${\cal O}_i^q$ are related to those of the colour-rearranged operators $\tilde{\cal O}_i^q$ by
\begin{equation}
    \langle \tilde {\cal O}_i^q \rangle_{\cal B} = - \tilde B_i^q  \langle  {\cal O}_i^q \rangle_{\cal B}\,, \qquad i = 1,2,
    \label{eq:Relation_O_tilde}
\end{equation}
where ${\cal B} = \{\Lambda_b, \Xi_b,\Omega_b\}$ and we have introduced the compact notation $\langle  {\cal O} \rangle_{\cal B} \equiv \langle {\cal B}| {\cal O}| {\cal B}\rangle/(2 M_{\cal B})$. The antisymmetry of the baryon wave function implies that, in the valence quark approximation, $\tilde B^q_i = 1$~\cite{Neubert:1996we}. Neglecting subleading $1/m_b$ and $SU(3)_F$-breaking corrections, we assume a universal parameter $\tilde B_i^q = \tilde B = 1$ defined at a typical hadronic scale $\mu_h \ll m_b$. We take $\mu_h = 1.5\mbox{~GeV}$ as central value and vary it within the range $1\mbox{~GeV} \leq \mu_h \leq 1.5\mbox{~GeV}$. 

The available information on the baryonic matrix elements of the colour-singlet operators is currently rather limited. Following Ref.~\cite{Gratrex:2023pfn}, we estimate them within the non-relativistic constituent quark model (NRCQM), in which the dimension-six four-quark operator matrix elements are expressed in terms of the baryon wave functions evaluated at the origin. For the $\Lambda_b$ baryon, one finds
\begin{equation}
\langle {\cal O}_1^q\rangle_{\Lambda_b} = - |\Psi^{\Lambda_b}(0)|^2\,,
\quad
\langle {\cal O}_2^q\rangle_{\Lambda_b} = \frac12 |\Psi^{\Lambda_b}(0)|^2\,,
\label{eq:4q_ME}
\end{equation}
with analogous relations holding for the remaining baryons, see, e.g.,~Ref.~\cite{Gratrex:2023pfn}. The wave functions at the origin can then be extracted from the observed hyperfine mass splittings~\cite{DeRujula:1975qlm}. Normalising the resulting expressions to the corresponding one for the $B$-meson wave function yields
\begin{equation}
\langle {\cal O}_1^q\rangle_{\Lambda_b} = - \frac{m_b^{\cal B} m_q^{\cal B}}{m_b^{\cal M} m_q^{\cal M}}  \frac{M_{\Sigma_b^*} - M_{\Sigma_b} }{M_{B^*} - M_B} \frac{F_B^2 (\mu_h)}{9}\,,
\label{eq:O1}
\end{equation}
where $m_{b,q}^{\cal B}$, $m_{b,q}^{\cal M}$ denote the constituent masses of the heavy and light quarks in the $b$ baryon and $B$ meson~\cite{Karliner:2014gca}, respectively, and $F_B$ is the HQET decay constant. 
Analogous expressions hold for the remaining baryons.

Eq.~\eqref{eq:O1} allows us to estimate the colour-singlet matrix elements. The HQET decay constant can be related, up to corrections of order $1/m_b$, to the QCD decay constant $f_B$, whose value is precisely determined by lattice QCD~\cite{FLAG:2024}.
However, following Ref.~\cite{Gratrex:2023pfn}, we express $F_B$ in terms of the QCD decay constant defined in the static limit $\hat f_{B}$. Lattice QCD determinations of the latter can be found in Ref.~\cite{Aoki:2014nga}.
Using this relation, we obtain the numerical values of the matrix element of the colour-singlet operator ${\cal O}_1^q$, for the $\Lambda_b, \Xi_b$, and $\Omega_b$ baryons, displayed in Tab.~\ref{tab:O1Numerics}. The corresponding results for the remaining four-quark operators are obtained using Eqs.~\eqref{eq:Relation_O_tilde} and \eqref{eq:4q_ME}. Finally, one-loop renormalisation group equations for the HQET operators~\cite{Neubert:1996we, Kirk:2017juj} are used to evolve these parameters from the hadronic scale $\mu_h$ to the $b$-quark scale $\mu_b \sim m_b$. 

Before proceeding with the discussion of the other non-perturbative inputs, it is worth emphasising that, in the case of the $\Lambda_b^0$ baryon, alternative determinations of the relevant hadronic matrix elements are available in the literature, based on HQET sum rules~\cite{Colangelo:1996ta,Wu:2026kzg} and QCD sum rules~\cite{Zhao:2021lzd,Wu:2026kzg}. It is therefore instructive to compare these results with our estimate obtained within the NRCQM. To this end, we introduce the following parametrisation
\begin{equation}
\langle \Lambda_b| {\cal O}_1^q | \Lambda_b \rangle = f_B^2 M_B  M_{\Lambda_b} L_1\,,
\label{eq:L1}
\end{equation}
where $L_1$ is a dimensionless hadronic parameter, while $M_B$ and $M_{\Lambda_b}$ denote the masses of the $B$ meson and $\Lambda_b$ baryon, respectively. The numerical values of $L_1$ obtained from the different approaches are collected in Tab.~\ref{tab:L1}. The NRCQM result quoted there corresponds to the value of the matrix element given in Tab.~\ref{tab:O1Numerics}. 
\begin{table}[t]
\renewcommand{\arraystretch}{1.9}
\centering
\begin{tabular}{c|c}
\hline
& $L_1$ \\
\hline
HQET sum rule '96~\cite{Colangelo:1996ta} & $- 0.033 \pm 0.017$\\
\hline
QCD sum rule '21~\cite{Zhao:2021lzd} & $- 0.143\pm 0.028$\\
\hline
QCD sum rule '26~\cite{Wu:2026kzg} & $- 0.119 \pm 0.041$\\
\hline
HQET limit of QCD sum rule '26~\cite{Wu:2026kzg} & $- 0.177 \pm 0.068$\\
\hline
NRCQM (This work) & $- 0.139 \pm 0.048$\\
\hline
\end{tabular}
\caption{Comparison of the values of the hadronic parameter $L_1$, defined in Eq.~\eqref{eq:L1}, obtained from HQET and QCD sum rules with the NRCQM result used in this work. The latter is derived from the value of the matrix element in Tab.~\ref{tab:O1Numerics}.}
\label{tab:L1}
\end{table}

We observe that, within uncertainties, all determinations are consistent with one another, with the exception of the older HQET sum rule result of Ref.~\cite{Colangelo:1996ta}. A detailed discussion of the possible origin of this discrepancy can be found in Ref.~\cite{Wu:2026kzg}. The good agreement between our NRCQM estimate and the more recent sum rule determinations strongly supports its use for the other baryons considered in this work. This is particularly important since, at present, no independent determinations of the corresponding matrix elements for the $\Xi_b$ and $\Omega_b$ baryons are available.

\begin{table*}[th]
\renewcommand{\arraystretch}{2.}
    \centering
    \begin{tabular}{l|c|c|c|c} 
      \hline
      & $\Lambda_b^0$ & $\Xi_b^{0,-}$  & $\Omega_b^-$ & $B^0_d$  
      \\ 
      \hline
      $[\mu_\pi^2 ({\cal B})]^{\rm kin}$ 
      & $0.483 \pm 0.046$ 
      & $0.515 \pm 0.052$ 
      & $0.574 \pm 0.080$ 
      & $0.454 \pm 0.043$
      \\
      \hline
      $\mu_G^2 ({\cal B})$  
      & 0 
      & 0 
      & $0.193 \pm 0.068$ 
      & $0.274 \pm 0.053$
      \\ 
      \hline
      $[\rho_D^3 ({\cal B})]^{\rm OS}$ 
      & $0.031 \pm 0.011$  
      & $0.037 \pm 0.010$ 
      & $0.050 \pm 0.018$ 
      & $0.030 \pm 0.003$
      \\ 
      $[\rho_D^3 ({\cal B})]^{\rm kin}$ 
      & $0.171^{+0.033}_{-0.024}$  
      & $0.177^{+0.033}_{-0.023}$ 
      & $0.190^{+0.036}_{-0.028}$ 
      & $0.170^{+0.031}_{-0.021}$
      \\ 
      \hline
    \end{tabular}
    \caption{Numerical values of dimension-five, in GeV$^2$, and dimension-six, in GeV$^3$, non-perturbative parameters used in our analysis. For the parameters defined in the kinetic scheme, the values quoted correspond to $\mu^{\rm cut} = 1$ GeV. The value in the on-shell scheme, corresponds to $\mu^{\rm cut} = 0$. The quoted uncertainties are obtained by adding in quadrature the parametric uncertainties and the estimated uncertainties associated with neglected higher-order power corrections as discussed in Ref.~\cite{Gratrex:2023pfn}.}
\label{tab:2q-papameters}
\end{table*}

We next consider the matrix element of the dimension-six two-quark operator, namely the Darwin operator. It is parametrised in terms of the Darwin parameter
\begin{equation}
   \rho^3_D({\cal B}) = \frac{1}{2M_{\cal B}} \langle {\cal B}| \bar b_v \left(i D_\mu\right) \left(i v\cdot D \right)
   \left(i D^\mu\right) b_v| {\cal B}\rangle\,,
\end{equation}
where $b_v$ is the rescaled heavy-quark field~\footnote{Differences between using $b_v$ and the HQET field $h_v$ first arise  at order $1/m_b^4$.}
\begin{equation}
b_v (x) = e^{i m_b v\cdot x} b(x)\,,
\end{equation}
$v^\mu$ denotes the $b$-baryon four-velocity and $D_\mu = \partial_\mu - i g_s A_\mu$ is the covariant derivative with respect to the background soft-gluon field. 

The relation between the Darwin parameter in the pole and kinetic schemes~\cite{Bigi:1996si, Czarnecki:1997sz} is given by
\begin{equation}
[\rho_D^3]^{\rm OS} = [\rho_D^3 (\mu^{\rm cut})]^{\rm kin} - \left[\rho_D^3 (\mu^{\rm cut})\right]_{\rm pert},   
\label{eq:rhoD3-pole-to-kin}
\end{equation}
where $\mu^{\rm cut}$ is  the Wilsonian cutoff, corresponding to a non-perturbative scale satisfying $\mu^{\rm cut}\ll m_b^{\rm kin}$, with $m_b^{\rm kin}$ being the value of the $b$-qaurk mass in the kinetic scheme. In the following, we take $\mu^{\rm cut} = 1 \GeV$ as central value and vary it within the interval $0.7\leq \mu^{\rm cut} \leq 1.3$ GeV. The second term on the r.h.s.\ of Eq.~\eqref{eq:rhoD3-pole-to-kin},  $[\rho_D^3 (\mu^{\rm cut})]_{\rm pert}$ denotes the corresponding perturbative contribution, which is known up to three-loop accuracy~\cite{Fael:2020iea, Fael:2020njb}.

Also in the case of the Darwin parameter, the available information is rather limited. In the absence of further input, either from direct calculations or from fits to semileptonic data, one may exploit equation-of-motion~(EOM) relations to relate the matrix element of the Darwin operator to those of the dimension-six four-quark operators. This has been recently done, for example, in Ref.~\cite{Egner:2024lay}, where the value of the Darwin parameter of the $B^0_s$ meson was obtained from the corresponding parameter for the $B_d^0$ meson, extracted from a fit of the semileptonic data \cite{Finauri:2023kte}, using EOM relations, see Ref.~\cite{Egner:2024lay} for more details. 

In the baryonic case, however, analogous semileptonic fits are not yet available due to the lack of sufficiently precise experimental data. We therefore rely entirely on EOM relations. Specifically, we use
\begin{align}
& 2 M_{\cal B} \, [\rho_D^3 ({\cal B})]^{\rm OS} = \frac{g_s^2}{24} 
\label{eq:rhoD3-eom} \\[1mm]
& \times \!\! \sum_{q = u, d, s} \!\! \langle {\cal B}| \left(- 3 {\cal O}_1^q + \tilde {\cal O}_1^q + 6 {\cal O}_2^q - 2 \tilde {\cal O}_2^q \right) |{\cal B} \rangle + {\cal O} \left(\frac{1}{m_b} \right) \,, 
\nonumber
\end{align}
where the colour-singlet and colour-rearranged operators, ${\cal O}^q_{1,2}$ and $\tilde {\cal O}^q_{1,2}$, are 
defined in Eqs.~\eqref{eq:O1q}-\eqref{eq:O2tq}.

Eqs.~\eqref{eq:rhoD3-pole-to-kin} and \eqref{eq:rhoD3-eom} allow us to determine the Darwin parameter in the pole and kinetic schemes, respectively, for all $b$ baryons, using as input the value of the strong coupling at the scale $\mu_b$ together with the dimension-six four-quark matrix elements listed in Tab.~\ref{tab:O1Numerics}. The resulting numerical values are collected in Tab.~\ref{tab:2q-papameters}. We note that, compared to Ref.~\cite{Gratrex:2023pfn}, the values of the Darwin parameter obtained from the EOM relations are now significantly larger. This difference arises because the perturbative contribution in Eq.~\eqref{eq:rhoD3-pole-to-kin} was not yet included in the analysis of Ref.~\cite{Gratrex:2023pfn}.

For consistency, as we consider ratios of $b$-baryon lifetimes to the $B_d^0$-meson lifetime, we also determine the Darwin parameter of the $B$ meson using the EOM relation. Specifically, we employ Eq.~\eqref{eq:rhoD3-pole-to-kin} using results up to three-loop accuracy for the perturbative contribution~\cite{Fael:2020iea, Fael:2020njb}. The corresponding value is shown in the last column of Tab.~\ref{tab:2q-papameters}. This is in very good agreement with the most recent determination of the Darwin parameter in the kinetic scheme extracted from a fit to semileptonic $B$ decays~\cite{Finauri:2023kte}, namely
\begin{equation}
\rho_D^3 (B^0_d; \mu^{\rm cut} = 1 \GeV) = (0.176 \pm 0.019) \GeV^3\,.
\end{equation}  

We also note that a recent analysis~\cite{Melic:2025tsr} derived constraints on the kinetic and Darwin parameters of the $\Lambda_b^0$ baryon using Small Velocity Sum Rules for the inclusive
semileptonic decay $\Lambda_b \to X_c \, e^- \bar \nu_e$.
Combining the resulting constraints with the estimate of $\rho_D^3$, obtained from EOM relations using the NRCQM results for the dimension-six four-quark matrix elements from Ref.~\cite{Gratrex:2023pfn}, the authors obtained $[\rho_D^3(\Lambda_b^0)]^{\rm kin} \sim 0.07$ GeV$^3$, at the cutoff scale $\mu^{\rm cut} = 0.75 {\rm GeV}$. This result is smaller than the value given in Tab.~\ref{tab:2q-papameters}. However, since the perturbative contribution entering Eq.~\eqref{eq:rhoD3-pole-to-kin} was not included in determination obtained in Ref.~\cite{Gratrex:2023pfn}, a direct comparison with our result is not straightforward.

Finally, we briefly discuss the matrix elements of the dimension-five two-quark operators, which are parametrised by the kinetic and chromomagnetic parameters
\begin{align}
   \mu^2_\pi({\cal B}) &= - \frac{1}{2M_{\cal B}} \langle {\cal B}| \bar b_v \left(i D_\mu\right) 
   \left(i D^\mu\right) b_v| {\cal B}\rangle\,,\\[1mm]
   \mu^2_G({\cal B}) &=  \frac{1}{2M_{\cal B}} \langle {\cal B}| \bar b_v \left(i D_\mu\right) 
   \left(i D_\mu\right) (-i \sigma^{\mu\nu})b_v| {\cal B}\rangle\,.
\end{align}
The chromomagnetic parameter can be extracted from spectroscopy relations, while the kinetic parameter is obtained from the heavy-quark expansion of the baryon masses. We refer the reader to Ref.~\cite{Gratrex:2023pfn} for the details of these determinations. The numerical values adopted in our analysis are collected in Tab.~\ref{tab:2q-papameters}. 

We note that, compared to the previous analysis of Ref.~\cite{Gratrex:2023pfn}, the values of $[\mu_\pi^2({\cal B})]^{\rm kin}$ have been updated to account for the more recent determinations of the corresponding kinetic parameters for the $B_d^0$~\cite{Finauri:2023kte} and $B_s^0$~\cite{Egner:2024lay} mesons.  

\section{Results}
\label{sec:Results}
\begin{table*}[th]
\renewcommand{\arraystretch}{2.3}
\centering
\begin{tabular}{c|c|c|c|c}
    \hline
    Observable & HQE-LO & HQE-NLO & HQE-NNLO & 
    Experimental value \\
    \hline
    $\Gamma (\Lambda_b^0)$ 
    & $0.716^{+0.118}_{-0.129} \psin$
    & $0.678^{+0.045}_{-0.062} \psin$
    & $0.673^{+0.017}_{-0.031} \psin$
    & $(0.683 \pm 0.004) \psin$
    \\
    \hline
    $\Gamma (\Xi_b^0)$ 
    & $0.712^{+0.118}_{-0.129} \psin$ 
    & $0.676^{+0.046}_{-0.063} \psin$
    & $0.670^{+0.021}_{-0.033} \psin$
    & $(0.679 \pm 0.007) \psin$ 
    \\
    \hline
    $\Gamma (\Xi_b^-)$ 
    & $0.665^{+0.111}_{-0.125} \psin$ 
    & $0.624^{+0.045}_{-0.061} \psin$
    & $0.618^{+0.020}_{-0.032} \psin$
    & $(0.635 \pm 0.009) \psin$
    \\
    \hline
    $\Gamma (\Omega_b^-)$ 
    & $0.625^{+0.111}_{-0.124} \psin$ 
    & $0.591^{+0.050}_{-0.064} \psin$
    & $0.585^{+0.029}_{-0.037} \psin$
    & $0.610^{+0.066}_{-0.054} \psin$ 
    \\
\hline
\end{tabular}
\caption{Comparison of the HQE predictions for the total decay widths of the $b$-baryons at LO, NLO, and NNLO in QCD, shown in the second, third, and fourth columns, respectively, with the corresponding experimental values given in the last column.
The experimental widths are obtained from the measured lifetimes~\cite{HFLAV} using the relation $\Gamma = 1/\tau$. 
}
\label{tab:Decay-width-HQE-vs-Data}
\end{table*}
In this section, we present our numerical analysis and provide updated predictions for the $\Lambda_b^0, \Xi_b^{0,-},$ and $\Omega_b^-$ baryon lifetimes and their lifetime ratios, as well as their ratios to the $B_d^0$ meson lifetime. 
In addition to the inputs listed in Tab.~\ref{tab:O1Numerics} and Tab.~\ref{tab:2q-papameters}, we adopt the remaining parameters, including the $\Delta B=1$ Wilson coefficients, Cabibbo-Kobayashi-Maskawa~(CKM) matrix elements, and quark masses, from the NNLO analysis of $B$-meson lifetimes in Ref.~\cite{Egner:2024lay}.

The lifetime is obtained from the relation $\tau = \Gamma^{-1}$, where the total width is computed using Eq.~\eqref{eq:HQE}. As for the lifetime ratio, we employ the following expression
\begin{equation}
\frac{\tau ({\cal B})}{\tau (H_b)} = 1 + \left[\Gamma (H_b) - \Gamma ({\cal B}) \right]^{\rm HQE} \tau ({\cal B})^{\rm exp}\,,
\label{eq:HQE-ratio}
\end{equation}
where $H_b$ denotes either a $b$-baryon or the $B_d^0$ meson. Note that we take the experimental value of the ${\cal B}$-baryon lifetime as an input due to the smaller uncertainty. Equivalently, the lifetime ratio can be computed entirely within the HQE, albeit with larger uncertainties.

\begin{figure}[th]
\centering
    \includegraphics[width=0.41\textwidth]{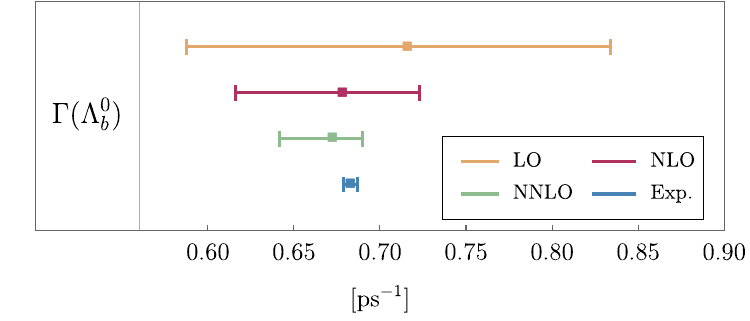} \\[2mm]
    \includegraphics[width=0.41\textwidth]{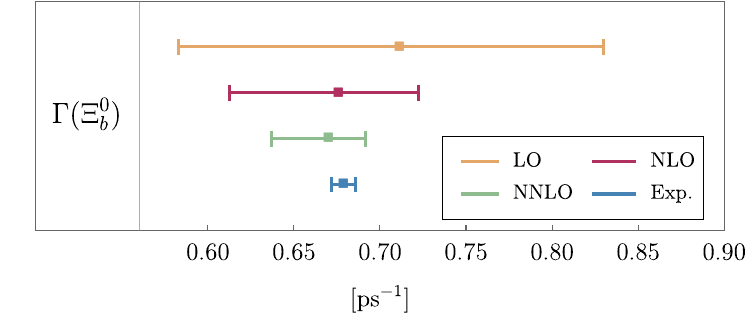} \\[2mm]
    \includegraphics[width=0.41\textwidth]{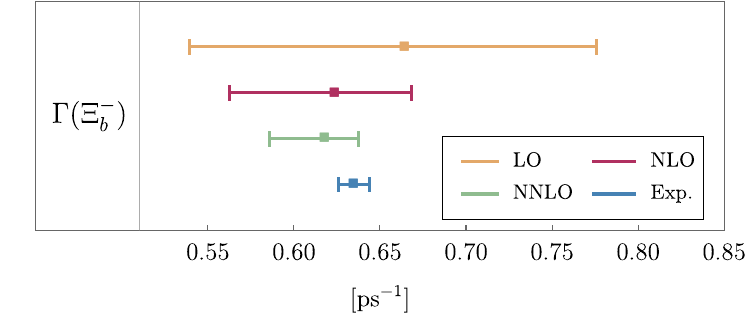} \\[2mm]
    \includegraphics[width=0.41\textwidth]{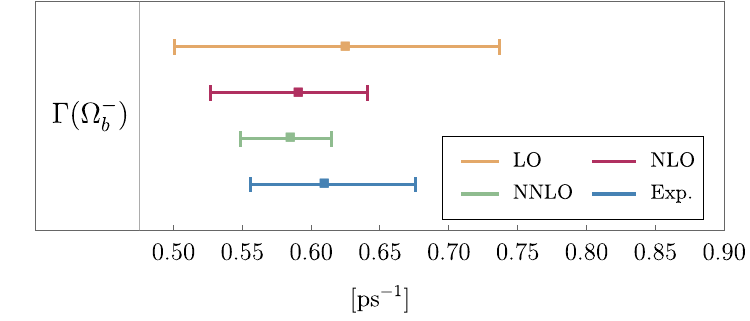}
\caption{Comparison of the HQE predictions for the total decay widths of the $\Lambda_b^0, \Xi_b^{0,-}$, and $\Omega_b^- $ baryons at LO (orange), NLO (magenta), and NNLO (green) with the corresponding experimental measurements (blue).}
\label{fig:Decay-widths-NNLO-NLO-LO}
\end{figure}

A comparison of the HQE predictions for the total widths of the $\Lambda_b^0, \Xi_b^{0,-}$, and $\Omega_b^-$ baryons at LO, NLO, and NNLO is shown in Fig.~\ref{fig:Decay-widths-NNLO-NLO-LO} and Tab.~\ref{tab:Decay-width-HQE-vs-Data}. 
The newly included ${\cal O}(\alpha_s^2)$ corrections to the free $b$-quark decay have only a minor effect on the central values. They do, however, significantly reduce the uncertainty associated with the variation of the renormalisation scales by roughly a factor of two, in agreement with the findings of Ref.~\cite{Egner:2024lay}. 

We also note that the NLO predictions for the total widths reported in Tab.~\ref{tab:Decay-width-HQE-vs-Data} differ slightly from those presented in the previous analysis of Ref.~\cite{Gratrex:2023pfn}. This difference is mainly due to updated input parameters, a different approach for the renormalisation scale settings as well as the correction of a typo that affected the renormalisation-group evolution of the dimension-six four-quark matrix elements.

\begin{figure}[th]
\centering
\includegraphics[scale=0.62]{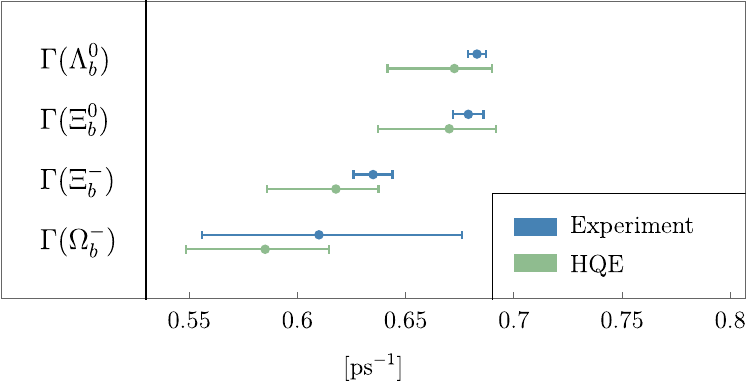}
\caption{Comparison of the HQE predictions for the total decay widths of the $\Lambda_b^0, \Xi_b^{0,-}$, and $\Omega_b^- $ baryons at NNLO (green) with the corresponding experimental measurements (blue).}
\label{fig:Decay-widths-HQE-vs-Data}
\end{figure}

A summary of the HQE predictions for the total decay widths at NNLO-QCD for all four baryons considered together with the experimental measurements is shown in Fig.~\ref{fig:Decay-widths-HQE-vs-Data}. 
Regarding the error budget, the residual renormalisation-scale dependence of the partonic contribution remains one of the dominant sources of theoretical uncertainty. Another significant contribution arises from the hadronic matrix elements, although their impact, being power-suppressed, is generally reduced for the total decay widths.

Overall, the theoretical predictions are in excellent agreement with the experimental measurements. The theoretical uncertainties are generally larger than the experimental ones, with the exception of the $\Omega_b^-$ baryon, for which the experimental uncertainty currently dominates. 

\begin{figure}[th]
     \centering
     \includegraphics[scale=0.62]{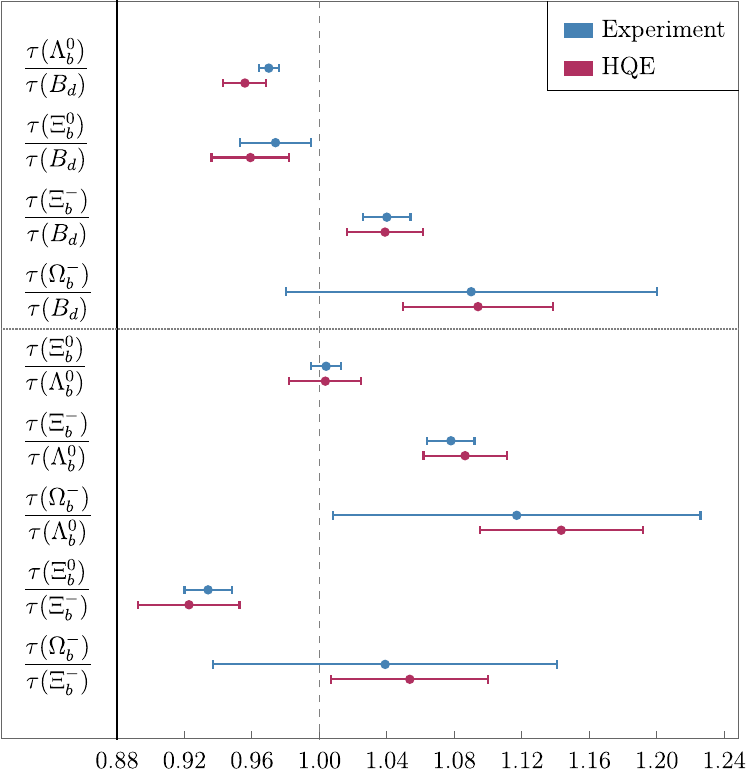}\\
     \caption{Comparison of the HQE predictions (magenta) for the lifetime ratios of the $\Lambda_b^0, \Xi_b^{0,-}$, and $\Omega_b^- $ baryons as well as their ratios with the $B_d^0$ meson with the corresponding experimental measurements (blue). As for the accuracy of the HQE predictions, see the description in the main text.}
\label{fig:Lifetime-ratios-HQE-vs-Data}
\end{figure}

Finally, our updated HQE predictions for the lifetime ratios of the $\Lambda_b^0, \Xi_b^{0, -}$ and $\Omega_b^-$ baryons as well as for their ratios relative to the $B_d^0$ lifetime are shown in Fig.~\ref{fig:Lifetime-ratios-HQE-vs-Data} and in Tab.~\ref{tab:Lifetime-ratios-HQE-vs-Data}.
We stress that, using the expression in Eq.~\eqref{eq:HQE-ratio}, the dependence on the partonic contribution to the lifetime ratios exactly cancels. This has several implications.

First, the results are not affected by the inclusion of ${\cal O}(\alpha_s^2)$ corrections to the free $b$-quark decay. In particular, while for the total width the perturbative accuracy of the partonic contribution largely determines the accuracy of the HQE prediction, since the remaining contributions are power suppressed, this is not the case in lifetime ratios, where a well-defined perturbative accuracy has not yet been achieved. In fact, while the dimension-five and dimension-six four-quark contributions are included at NLO-QCD, the dimension-six two-quark contribution is implemented only at LO. We therefore refrain from quoting a definite perturbative order in QCD for our results.

Second, the lifetime ratios are driven by the size of the power corrections and are highly sensitive to them. The inclusion of ${\cal O}(\alpha_s)$ corrections to dimension-five operators, namely the kinetic and chromomagnetic operators, leads to a visible shift, particularly for the latter, and always in the direction of the experimental determinations, thereby improving the overall agreement between HQE predictions and data. We note that, despite the inclusion of these corrections and the use of updated input for hadronic parameters, in particular $[\rho_D^3]^{\rm kin}$, there is only a small difference with respect to the values of the previous study~\cite{Gratrex:2023pfn}. This is because of a partial compensation arising from a typo in the running of the dimension-six four-quark matrix elements.

The uncertainty budget is dominated by the uncertainties of the hadronic matrix elements as well as the renormalisation scale dependence. Compared to the previous study~\cite{Gratrex:2023pfn}, the quoted errors are slightly different. On the one hand, this is due to updated input, for instance the additional renormalisation-scale dependence in the $\overline{\rm MS}$ scheme for the charm quark, now varied independently compared to before, as well as the $\mu^{\rm cut}$ variation. On the other hand, we now take into account correlations between dimension-six four-quark matrix elements.

Overall, we find excellent agreement for all observables considered. The theoretical uncertainties are comparable to the experimental ones in most cases, and significantly smaller for ratios involving the $\Omega_b^-$ baryon. Note that for some ratios with respect to the $B_d^0$ lifetime, no experimental value is available. In these cases, we take the ratio of the experimental measurement of the baryon lifetime to $\tau(B_d^0) = (1.517 \pm 0.004) \mathrm{ps}$~\cite{HFLAV}. Hence, possible experimental correlations are not taken into account.
We also note that the HQE prediction for $\tau(\Omega_b^-)/\tau(\Xi_b^-)$ is presented here for the first time.

\begin{table}[th]
\renewcommand{\arraystretch}{2.3}
\centering
\begin{tabular}{c|c|c}
    \hline
    Observable &  HQE & 
    Experimental value \\
\hline    
    $\tau (\Lambda_b^0)/ \tau(B_d^0)$  
    & $0.956 \pm 0.013$
    & $0.970 \pm 0.006$ \cite{HFLAV} 
    \\
    \hline
    $\tau (\Xi_b^0)/ \tau(B_d^0)$ 
    & $0.959 \pm 0.023$
    & $0.974 \pm 0.021 \, {^\diamond}$ \cite{HFLAV}  
    \\
    \hline
    $\tau (\Xi_b^-)/ \tau(B_d^0)$ 
    & $1.039 \pm 0.022$
    & $1.040 \pm 0.014 \, {^\diamond}$ \cite{HFLAV} 
    \\
    \hline
    $\tau (\Omega_b^-)/ \tau(B_d^0)$  
    & $1.094 \pm 0.045$
    & $1.09 \pm 0.11$ \cite{HFLAV} 
    \\
\hline
    $\tau (\Xi_b^0)/ \tau (\Lambda_b^0)$ 
    & $1.004 \pm 0.021$
    & $1.004 \pm 0.009$ \cite{LHCb:2025zft}  
    \\
    \hline
    $\tau (\Xi_b^-)/ \tau (\Lambda_b^0)$ 
    & $1.086 \pm 0.025$
    & $1.078 \pm 0.014$ \cite{LHCb:2024lao}  
    \\
    \hline
    $\tau (\Omega_b^-)/ \tau (\Lambda_b^0)$ 
    & $1.143 \pm 0.049$
    & $1.117 \pm 0.109 \, {^\diamond}$ \cite{HFLAV}
    \\
    \hline
    $\tau (\Xi_b^0)/ \tau (\Xi_b^-)$ 
    & $0.923 \pm 0.030$
    & $0.934 \pm 0.014$ \cite{HFLAV} 
    \\
    \hline
    $\tau (\Omega_b^-)/\tau (\Xi_b^-)$ 
    & $1.054 \pm 0.046$
    & $1.039 \pm 0.102 \, {^\diamond}$ \cite{HFLAV}
    \\
    \hline
\end{tabular}
\caption{Comparison of the HQE predictions for the lifetime ratios of the $\Lambda_b^0, \Xi_b^{0,-}$, and $\Omega_b^- $ baryons as well as their ratios with the $B_d^0$ meson  (second column) with the corresponding experimental determinations (third column). 
Values marked by ${^\diamond}$ are obtained dividing the corresponding experimental value for the baryon lifetime by $\tau (B_d^0) = (1.517 \pm 0.004) \ps$~\cite{HFLAV}. Hence, possible experimental correlations are not taken into account. 
}
\label{tab:Lifetime-ratios-HQE-vs-Data}
\end{table}

\section{Conclusion and outlook}
\label{sec:Conclusion}
In this paper we have performed an updated analysis within the HQE of the $\Lambda_b^0, \Xi_b^{0,-}$, and $\Omega_b^-$ baryon lifetimes and lifetime ratios, as well as their ratios with respect to the $B_d^0$ lifetime. 

Compared to previous studies, we now include ${\cal O}(\alpha_s^2)$ corrections to the free $b$-quark decay rate and ${\cal O}(\alpha_s)$ corrections to the coefficient of the kinetic and chromomagnetic operators, which enter at order $1/m_b^2$. Furthermore, at order $1/m_b^3$, we improve the determination of the Darwin parameter in the kinetic scheme, by means of EOM relations, consistently including perturbative contributions up to three-loop accuracy. 

For the total widths, we find that the inclusion of ${\cal O}(\alpha_s^2)$ corrections leads to a significant reduction of the theoretical uncertainties, by roughly a factor two. Our predictions are in excellent agreements with the experimental measurements and, in the case of the $\Omega_b^-$ lifetime, achieve a precision which is better than the corresponding experimental determination. 

On the other hand, the inclusion of  ${\cal O}(\alpha_s^2)$ corrections does not affect our predictions for lifetime ratios, which are only sensitive to subleading power corrections. In this case, the ${\cal O}(\alpha_s)$ corrections at dimension five induce a visible shift in the central values, particularly through the chromomagnetic operator, and lead to an improved overall agreement with experimental data. We find remarkably good agreement for all lifetime ratios considered, both among $b$-baryons and between $b$-baryons and the $B_d^0$ meson. In these cleaner observables, the theoretical uncertainties are generally comparable to or smaller than the experimental ones.

Overall, our results provide a stringent and consistent test of the HQE framework in the baryon sector and further support its applicability to inclusive $b$-hadron decay widths at the current level of experimental and theoretical precision. Further progress will require several key improvements both on the perturbative and non-perturbative side. Specifically:
\begin{itemize}
\item[$\diamond$] Achieve full NLO-QCD accuracy for the coefficient of the Darwin operator by computing $\alpha_s$ corrections induced by non-leptonic $b$-quark decays at order $1/m_b^3$.

\item[$\diamond$] Achieve NNLO-QCD accuracy in the dimension-six four-quark operator contribution, extending the recent results of Ref.~\cite{Moretti:2026waq}. 
This requires either computing ${\cal O}(\alpha_s^2)$ corrections for operators defined in the HQET limit, or consistently accounting for operator mixing between dimension-six four-quark operators and dimension-three two-quark operators defined in QCD. In this context, progress in lattice QCD will be essential.

\item[$\diamond$] Improve the determination of the matrix elements of two-quark operators by extracting them from fits to semileptonic $b$-baryon decays. This will require more precise experimental data.

\item[$\diamond$] Improve the determination of the dimension-six four-quark matrix elements through a dedicated first-principles lattice QCD calculation. While there has recently been significant progress in the meson sector regarding the determination of four-quark matrix elements using HQET sum rules \cite{Black:2024bus,King:2021jsq,Kirk:2017juj} and lattice QCD \cite{Black:2026rbz,Black:2026dzp}, the baryon sector remains considerably less explored.
\end{itemize}

\section*{Acknowledgments}
We are grateful to Steven Roy Blusk for discussions on the experimental status and for useful correspondence. We also thank Lovro Dulibi\'c for helping us identifying a typo in the running of the dimension-six four-quark matrix elements used in Ref.~\cite{Gratrex:2023pfn}.
The research of AL is supported by the Deutsche Forschungsgemeinschaft (DFG, German Research
Foundation) under Germany’s Excellence Strategy – EXC 3107 –
Project-ID 533766364 and under grant 396021762-TRR 257.
The research of MLP is funded by the European Union's Horizon Europe Research and Innovation Programme under the Marie Sk{\l}odowska-Curie grant agreement No.~101204923.
AR acknowledges the support by the
Deutsche Forschungsgemeinschaft (DFG, German Research Foundation) - project number 541305755.

\bibliography{References}

\end{document}